\documentclass[prd, 
showpacs,preprintnumbers,amsmath,amssymb]{revtex4}

\usepackage{hyperref}

\DeclareMathOperator{\ch}{\cosh}
\DeclareMathOperator{\sh}{\sinh}

\newcommand{\dd}{\,\mathrm{d}}

\newcommand{\real}{\mathbb {R}}
\newcommand{\zahl}{\mathbb {Z}}
\newcommand{\eps}{\varepsilon}

\newcommand{\vphi}{\varphi}

\newcommand{\lapl}{\Delta}

\newcommand{\ee}{\end{equation}}
\newcommand{\nn}{\nonumber}
\newcommand{\eq}[1]{(\ref{#1})}
\newcommand{\be}{\begin{equation}}
\newcommand{\bel}[1]{\begin{equation}\label{#1}}

\begin{document}

\title{Two black hole initial data}

\author{Szymon \L\c{e}ski}

\affiliation{Centrum Fizyki Teoretycznej\\ Polska Akademia Nauk\\
Al.~Lotnik\'ow 32/46, Warsaw, Poland}

\email{szleski@cft.edu.pl}
\begin{abstract} The
Misner initial data are a standard example of time-symmetric
initial data with two apparent horizons. Compact formulae
describing such data are presented in the cases of equal or
nonequal masses (i.e. isometric or non-isometric horizons). The
interaction energy in the ``Schwarzschild + test particle'' limit
of the Misner data is analyzed.
\end{abstract}

\pacs{04.20.Ex, 04.70.Bw}


\maketitle

\section{Introduction}In \cite{mis:wic} Misner proposed two-body wormhole
initial data and in \cite{mis:mig} he used the method of images to
describe the time-symmetric initial data for an arbitrary number
$N$ of particles (see also \cite{giu:cts} for a review of
time-symmetric initial data). Such data may be viewed as a
collection of Einstein-Rosen bridges connecting two isometric
sheets. If we restrict ourselves to just one of them, we can view
such initial data as a three-dimensional metric on a manifold with
internal boundary. The boundary consists of $N$ minimal surfaces
representing surfaces of the ``black holes'' corresponding to the
particles. In this paper I am concerned with the $N=2$ case only.
If the masses of black holes are equal (i.e. when the horizons are
isometric) then there is a well-known formula for the
three-metric:
\bel{Misg} g = \Phi^4 (\dd \mu^2 + \dd \eta^2 + \sin^2\eta \dd
\vphi^2)\;,\ee where
\bel{Misphi}\Phi=\sum_{n\in\zahl}
\frac{\sqrt{d}}{\sqrt{\ch(\mu+2n\mu_0) - \cos\eta}}\;.\ee Here
$(\mu,\eta,\vphi)$ are the bispherical coordinates defined by
\begin{eqnarray}
 \nn x &=& \cos\vphi\frac{\sin\eta}{\ch\mu - \cos\eta}\;,\\
 \nn y &=& \sin\vphi \frac{\sin\eta}{\ch\mu - \cos\eta}\;,\\
 z &=& \frac{\sh\mu}{\ch\mu - \cos\eta}\label{zet}\;.
\end{eqnarray}
There are two parameters in \eq{Misphi}: $d$ and $\mu_0$. The
$\mu$ coordinate ranges from
 $-\mu_0$ to $\mu_0$. The $\mu=\pm\mu_0$ surfaces are
minimal, which translates to the Neumann boundary condition for
$\Phi$: $\frac{\partial}{\partial\mu} \Phi|_{\mu=\pm\mu_0}=0$. The
conformal factor $\Phi$ satisfies an elliptic equation
\bel{eqphi} \sin\eta\left(\lapl -\frac14\right) \Phi = -4\pi\sqrt
d\delta_0\;,\ee where $\lapl$ is the Beltrami-Laplace operator
associated with the metric $\dd \mu^2 + \dd \eta^2 + \sin^2\eta
\dd \vphi^2$ and $\sin\eta$ is the volume element.

It is obvious that the metric $g$ is conformally related to the
Euclidean metric on $\real^3$ minus two balls which  correspond to
$|\mu|>\mu_0$. This transformation reads:
\bel{psimetric}
g = \Psi^4 d^2 (\dd x^2 + \dd y^2 + \dd z^2)\;,\ee where the
conformal factor  $\Psi$ is related to $\Phi$ by the formula $\Psi
= d^{-1/2}\Phi\sqrt{\ch\mu - \cos\eta}$. The $\sqrt{d}$ factor is
extracted from $\Phi$ in order to normalize the value of $\Psi$ at
infinity to $1$. Now the parameter $2d$ has the interpretation of
the Euclidean distance (with respect to the metric $d^2 (\dd x^2 +
\dd y^2 + \dd z^2)$) between the points corresponding to
$\mu=\pm\infty$ ($z=\pm1$).

To generalize the metric $g$ to the case of nonequal masses one
takes $\mu\in[-a,b]$ with $a,b>0$, imposes the Neumann boundary
conditions and solves equation \eq{eqphi} for $\Phi$. Such metrics
are well known and used to construct initial data. There are
effective formulae which give the conformal factor, see for
example \cite{coo:ida}. However, it seems that no closedform
formula has been published, for example the formulae from
\cite{mis:mig} include series of operators of inversion with
respect to spheres in Euclidean space and the formulae in
\cite{coo:ida} are recursive. Below I present a compact form of
the metric. First I derive a new version of formula \eq{Misphi} in
the equal-mass case, see \eq{PhiSing} and \eq{PhiNonSing}. Then I
present the corresponding formulae \eq{PhiSing2} and
\eq{PhiNoSing2} in the case of nonequal masses. I also use
\eq{PhiNoSing2} to calculate the interaction energy in the
test-body limit of the Misner initial data, that means when one
mass is much smaller than the other.

\section{Main result}Let us begin with
$$\frac{1}{|\vec{r}-\vec{r}{\,}^\prime|}=\sum_{l=0}^{\infty} \frac{r^l_{<}}{r^{l+1}_{>}}
P_l(\cos \beta)\;,$$ where $r_{<} := \min(|\vec{r}|,
|\vec{r}{\,}^\prime|)$, $r_{>} := \max(|\vec{r}|,
|\vec{r}{\,}^\prime|)$, $\beta$ --- the angle between $\vec{r}$,
$\vec{r}{\,}^\prime$. Taking $\vec{r}=\vec{e}_z$ and
$|\vec{r}{\,}^\prime| = \exp(\alpha)$, we get
$$|\vec{e}_z - \vec{r}{\,}^\prime| = \sqrt 2 e^{\alpha/2}
\sqrt{\ch\alpha - \cos\eta}\;,$$ where $\eta$ is a spherical angle
of the vector $\vec{r}{\,}^\prime$. Hence we have the following
formula for terms in \eq{Misphi}:
\bel{glwzor}\frac{1}{\sqrt{\ch\alpha-\cos\eta}} = \sqrt 2 \sum_{l=0}^\infty
e^{-|\alpha|(l+1/2)}P_l(\cos\eta)\;,\ee where $\alpha\neq 0.$ If
we expand the summands in \eq{Misphi} using \eq{glwzor} and sum
the $n$-indexed series then we arrive at the conformal factor
\eq{Misphi} for the equal-mass case rewritten as:
\bel{PhiSing}\Phi = \sqrt{2d} \sum_{l=0}^\infty
P_l(\cos\eta)\frac
{\ch[(\mu_0-|\mu|)(l+1/2)]}{\sh[\mu_0(l+1/2)]}\;,\ee the formula
being valid for $\mu\neq0$. We can also extract the singular term
$\frac{\sqrt d}{\sqrt{\ch\mu-\cos\eta}}$ and get the following
version of \eq{PhiSing}:
\begin{eqnarray}\label{PhiNonSing}\lefteqn{\Phi =
\frac{\sqrt d }{\sqrt{\ch\mu-\cos\eta}}} \\&& + \sqrt{2d}
\sum_{l=0}^\infty P_l(\cos\eta)e^{-\mu_0(l+1/2)}\frac
{\ch[\mu(l+1/2)]}{\sh[\mu_0(l+1/2)]}\;,\nn\end{eqnarray}with the
series being uniformly convergent for $\mu\in[-\mu_0, \mu_0]$.

In the case of non-equal masses the conformal factor is given by
the following formula: \bel{MisPhi2} \Phi = \frac{\sqrt d
}{\sqrt{\ch\mu-\cos\eta}} + \sqrt d \sum_{n=1}^{\infty}\left(
Q_{-2na-2(n-1)b} + Q_{-2na-2nb}+ Q_{2(n-1)a+2nb}+
Q_{2na+2nb}\right)\;,\ee where
$$Q_\alpha := \frac{1}{\sqrt{\ch\alpha-\cos\eta}}\;.$$ Simple symmetry
argument shows that such $\Phi$ indeed satisfies the Neumann
boundary conditions at $\mu=-a$, $\mu=b$. As before we use
\eq{glwzor} and perform summation over $n$ to get $\Phi$ in the
following compact form:
\bel{PhiSing2} \Phi = \sqrt{2d}\sum_{l=0}^\infty P_l(\cos\eta) \frac
{\ch[(a-b+\mu)(l+1/2)]+\ch[(a+b-|\mu|)(l+1/2)]}{\sh[(a+b)(l+1/2)]}\;,\ee
or, if we extract the singular term,
\begin{eqnarray}\label{PhiNoSing2}\lefteqn{\Phi = \frac{\sqrt
d}{\sqrt{\ch\mu-\cos\eta}} +}\\&& \sqrt{2d} \sum_{l=0}^\infty
P_l(\cos\eta)\frac{e^{-\mu(l+1/2)}(e^{2b(l+1/2)}+1)+e^{\mu(l+1/2)}(e^{2a(l+1/2)}+1)}
{e^{2(a+b)(l+1/2)}-1}\;.\nn\end{eqnarray} Again formula
\eq{PhiSing2} is valid for $\mu\neq0$ and the series in
\eq{PhiNoSing2} converges uniformly for $\mu\in[-a,b]$. If we
substitute $a=b=\mu_0$ then \eq{PhiSing2} and \eq{PhiNoSing2}
reduce to \eq{PhiSing} and \eq{PhiNonSing}, respectively.

The main result here are formulae \eq{PhiSing2} and
\eq{PhiNoSing2}. It is worth noting that the conformal factor is
expanded in a series of orthogonal polynomials, which is a
desirable feature for numerical treatment.

\section{Schwarzschild + Test-body limit} Let us now analyze the
``Schwarzschild + test body'' limit of Misner initial data. First,
we observe that if we pass to the limit $b\to \infty$ in
\eq{PhiNoSing2} then we get Schwarzschild initial data in
bispherical coordinates. The $\mu=-a$ sphere represents the
minimal surface in Schwarzschild initial data and $\mu=\infty$
becomes a regular point. The mass of such data is
\bel{mzero}m=\frac{2d}{\sinh a}\;.\ee Second, we treat $\eps :=
\exp(-b)$ as a small parameter, that means $\eps \ll \exp({-a})$,
and expand \eq{PhiNoSing2} in $\eps$. We get the following formula
for the conformal factor:
\begin{eqnarray}\label{tb}\lefteqn{\Phi =
\frac{\sqrt d}{\sqrt{\cosh\mu-\cos\eta}}}\\&& +
\sqrt{2d}\sum_{l=0}^\infty \frac{e^{-\mu(l+1/2)}}{e^{2a(l+1/2)}}
P_l(\cos\eta) + (1+e^{-a})(e^{\mu/2}+e^{-\mu/2-a})\eps +
O(\eps^2)\;.\nn
\end{eqnarray}
Let us denote by $m_1$ and $m_2$ the masses of the $\mu=-a$ and
the $\mu=b$ surfaces respectively. By ``mass of a surface'' I mean
here the square root of its area divided by $\sqrt{16\pi}$. Using
\eq{tb} we find the individual masses of the surfaces and the ADM
mass $M$ of the whole system:
\bel{masa1} m_1 = \frac{2d}{\sinh a} + 8d(e^{-a}+e^{-2a})\eps
+O(\eps^2)\;,\ee
\be m_2 = 4d(1+e^{-a})^2\eps +O(\eps^2)\;,\ee
\be M = m_1 + m_2 - 8d(e^{-a}+e^{-2a})\eps
+O(\eps^2)\;.\ee In the Newtonian gravity the interaction energy
equals $\mathring{E}=-\frac{m_1m_2}{\mathrm{distance}}$ and for
the Misner data the interaction energy should converge to this
value in the Newtonian limit, that means when the distance
increases to infinity. We would like to write the interaction
energy in the form $E = -\frac{m_1m_2}{\mathrm{distance}}\gamma$,
where $\gamma$ is a function of distance and converges to $1$ as
the distance goes to infinity. The two most obvious choices for
the distance parameter are $D:=2d$ and the length of the
$\eta=\pi$ geodesic connecting two horizons which we denote by
$L$. The distance $D$ is quite close to the distance studied by
Brill and Lindquist \cite{bl:ieg}, in the sense that it is the
Euclidean distance between two points lying inside minimal
surfaces. In this case we get the following formula for the
interaction energy:
\bel{int1} M-m_1-m_2 = -\frac{m_1m_2}{D} \left(
1-\frac{\frac{m_1}{D}}{1+\sqrt{1+\left(\frac{m_1}{D}\right)^2}}\right)
+ O(\eps^2)\;.\ee Passing now to the second case, we can calculate
the geodesic distance $L$ as the integral
$$L=d\int_{z_1}^{1} (\Psi(x=0, y=0, z))^2 \dd z
+  O(\eps)\;,$$ where $z_1=-\frac{\sh a}{\ch a + 1}$ is the $z$
coordinate \eq{zet} of the point $(\mu=-a, \eta=\pi)$. The result
is
$$ L = 2d\left[\frac12 - \frac{z_1}{2} + \frac{m_1}{2d}\log\frac{1-z_0}{z_1-z_0}
+
\frac{m_1^2}{8d^2}\left(\frac{1}{z_1-z_0}-\frac{1}{1-z_0}\right)\right]
+ O(\eps)\;$$ with $z_0 = - \frac{\ch a}{\sh a}$ being the $z$
coordinate of the center of the removed ball corresponding to the
$\mu=-a$ minimal surface. As the result we get the interaction
energy equal to
\bel{int2}M-m_1-m_2 = -\frac{m_1m_2}{L} \gamma_L + O(\eps^2)\;.\ee
The factor $$\gamma_L := \frac{L}{2d}\left(
1-\frac{\frac{m_1}{2d}}{1+\sqrt{1+\left(\frac{m_1}{2d}\right)^2}}\right)$$
can be expressed in zeroth order in $\eps$ (using \eq{masa1}) as a
rather complicated function of a single parameter
$\frac{2d}{m_1}$. In the $d\to\infty$ limit ($L\to\infty$) it
converges to $1$, giving the correct Newtonian limit of \eq{int2}.

\section*{ACKNOWLEDGMENTS} This research was supported by the Polish
Research Council grant KBN 2 P03B 073 24 and by the Erwin
Schr\"odinger Institute.

\end{document}